\documentclass[a4paper,UKenglish]{lipics}
 
\usepackage{microtype}

\title{Consistency in Distributed Data Stores}
\titlerunning{Consistency in Replicated Distributed Databases} 

\author{Mohammad Roohitavaf}
\affil{Department of Computer Science and Engineering\\
  Michigan State University\\
  East Lansing, MI, USA\\
  \texttt{roohitav@msu.edu}}
\authorrunning{M. Roohitavaf} 

\newtheorem{observation}{\textbf{Observation}}


\begin{document}

\maketitle

\begin{abstract}
This paper focuses on the problem of consistency in distributed data stores. We define strong consistency model which provides a simple semantics for application programmers, but impossible to achieve with availability and partition-tolerance. We also define weaker consistency models including causal and eventual consistency. We review COPS and GentleRain as two causally consistent data stores as well as Dynamo as an eventually consistent data store. We provide insights about scenarios where each of these methods is suitable, and some future research directions. 
 \end{abstract}
\section{Introduction}
Distributed data stores are among the fundamental building blocks of today's Internet services. Service providers usually replicate their data on different nodes world-wide to achieve higher performance and availability. However, when we use this approach, the consistency between replicas becomes a concern. In an ideal situation, all replicas always represent exactly the same data. In other words, any update to any data item instantaneously becomes visible in all replicas. This model of consistency is called \textit{strong consistency}. Strong consistency provides a simple semantics for programmers who want to use the distributed data store for their applications. Unfortunately, we cannot achieve strong consistency without sacrificing the availability when we may have network partitions. In particular, in case of network partitions, to maintain the consistency between different copies of data items, we should make updates unavailable. The CAP theorem implies that from strong consistency, availability, and partition-tolerance, we can have only two of them.

At the other end, the weakest consistency model is called \textit{eventual consistency}. In this consistency model, as the name suggests, the only guarantee is that replicas become consistent "eventually". We can implement always-available services under this consistency model. However, it may lead to undesired situations. For instance, suppose in a social network, Alice changes the status of an album to permission-required. Then, she uploads a personal photo to the album. Under eventual consistency model, the photo may become visible in a remote node before album status. In that case, any user querying the remote node can see Alice's personal photo without her permission until the new status of the album reaches the remote node. Despite such anomalies, some distributed data stores use eventual consistency. One example is Dynamo \cite{Dy} which is used by Amazon.com to manage the state of some of its services with very high availability requirements.

\textit{Causal consistency} is an intermediate consistency model. It is weaker than strong consistency, but stronger than eventual consistency. In causal consistency, we keep track of the dependencies between events in the system, to avoid certain anomalies. In particular, causal dependency requires that the effect of an event can be visible only when the effects of its dependencies are visible. For example, any event by a client depends on all previous events by that client. Thus, in our example, the event of uploading the photo depends on the event of changing the album status to permission-required. Therefore, nobody can view Alice's photo before finding her album as permission-required. Causal consistency is achievable with availability and partition-tolerance. COPS \cite{COPS} and GentleRain \cite{Gen} are two causally consistent distributed data stores. These systems use two different techniques. COPS uses explicit dependency check through dependency check messages. On the other hand, GentleRain uses implicit dependency check through physical timestamps.

In this paper, we describe strong consistency model, and discuss impossibility of achieving strong consistency with availability and partition-tolerance. We also review other consistency models weaker than strong consistency. We review COPS and GentleRain as two implementations of causal consistency, and Dynamo as an implementation of eventual consistency. We compare and contrast these implementations from different aspects. Our goal in this comparison is to provide insights about which technique is best for which scenario.

This paper is organized as follows: Section \ref{sec:CMs} discuses consistency models. Section \ref{sec:CMD} reviews COPS and GentleRains. Section \ref{sec:Dy} reviews Dynamo. Section \ref{sec:D_CG_com} compare Dynamo with COPS and GentleRain. Finally, Section \ref{sec:Con} concludes the paper and provides some future research directions.
\section{Consistency Models}
\label{sec:CMs}
\subsection{Preliminaries}

In order to provide a formal definition for strong consistency, we use definitions provided in \cite{Lin}. We define the \textit{system} to be a finite set of \textit{processes} (or nodes) which interact using a shared memory. The data store consists of a finite set of \textit{objects}. A Process can perform different \textit{operations} on data objects. In the context of data stores, we have two basic operations on data objects namely $PUT$ and $GET$. Each operation has a \textit{call} event and a \textit{response} event. We denote the call event of operation $op$ by $call.op$, and similarly the response event by $res.op$. A response event \textit{matches} a call event if their object names agree and their process names agree.

We model an execution of a concurrent system by a \textit{history}, which is a finite sequence of operation call and response events. A history $H$ is \textit{sequential} if: 
\begin{enumerate}

\item The first event of $H$ is a call event, 
\item Each call, except possibly the last, is immediately followed by a matching response. 
\end{enumerate}

A \textit{process subhistory}, $H|p$ ($H$ at $p$), of history $H$ is the subsequence of all events in $H$ whose process names are $p$. An \textit{object subhistory}, $H|o$,  is  similarly defined for an object $o$. Two histories $H$ and $H'$ are \textit{equivalent} if for every process $p$, $H|p =  H'|p$. If $H$ is a history, $complete(H)$ is the maximal subsequence of $H$ consisting only of call events and matching response events.

A set $S$ of histories is \textit{prefix-closed} if, whenever $H$ is in $S$, every prefix of $H$ is also in $S$. A \textit{single-object} history is one in which all events are associated with the same object.  A \textit{sequential specification} for an object is a prefix-closed set of single-object sequential histories for that object. A sequential history $H$ is \textit{legal} if each object subhistory $H|o$ belongs to the sequential specification for $o$. In the context of distributed data stores, the sequential specification for each object is the set of all single-object sequential histories for that object in which any $GET$ operation returns the value written by the most recent $PUT$ operation. However, object in different systems may have different sequential specifications. For instance, the sequential specification for a queue object follows FIFO order for enqueue and dequeue operations.
  
\subsection{Strong Consistency}
Strong consistency is another name for the notion of linearizability defined in \cite{Lin}. Linearizability provides a single image of the data across the system. That is equivalent to the scenario where the whole data is stored in a single node which responds to requests one at a time. Therefore, at any given time there is no difference between data seen by different nodes in the system. We formally define linearizable history as follows:

\begin{definition} [Linearizable History] 
\label{def:lin}
A history $H$ is linearizable if it can be extended (by appending zero or more response events) to some history $H'$ such that: 
\begin{itemize}
\item complete($H'$) is equivalent to some legal sequential history $S$, and 
\item whenever the $res.op_1$ precedes the $call.op_2$ in $H$, then $op_1$ precedes $op_2$ in $S$. 
\end{itemize}
\end{definition}

In order to show that history $H$ is linearizable, it is enough to provide a legal sequential history $H'$ equivalent to $H$ that satisfies both conditions of Definition \ref{def:lin}. A system is linearizable if all of its possible histories are linearizable. 

The second condition in Definition \ref{def:lin} captures the real-time precedence ordering of operations in $H$. If we remove this condition, we have the definition of the \textit{sequentially consistent} \cite{Seq} history. In both linearizability and sequential consistency, all processes should observe the same order of events. In linearizability this order respects real-time precedence ordering, while sequential consistency may not.

The CAP theorem \cite{CAP} shows that achieving strong consistency with availability and partition-tolerance is impossible. Availability means every request must result in an response. (Note that, failure is not an acceptable response.) Partition means network may lose arbitrary many messages sent from one node to another. When we say the system provides partition-tolerance and strong consistency, we mean system guarantees strong consistency while some messages may be lost. Similarly, when we say system provides partition-tolerance and availability, we mean every request must result in a response, while some messages may be lost.

Based on the definition of availability and partition-tolerance, CAP theorem is as follows: 

\begin{theorem}
\label{the:CAP}
It is impossible in the asynchronous network model to implement a read/write data object that guarantees the following properties: 
\begin{itemize}
\item  Availability
\item Strong consistency
\end{itemize}
In all fair executions (including those in which messages are lost). 
\end{theorem}

The proof of Theorem \ref{the:CAP} is provided in \cite{CAP}. Although CAP theorem is about strong consistency, we can show that achieving sequential consistency with availability and partition-tolerance is impossible as well.

Since providing strong consistency requires unavailability in case of network partitions, many practical systems consider weaker consistency models. We define some of them in the following section. 

\subsection{Weaker Consistency Models}
The weakest consistency model is eventual consistency. Under this consistency model, the storage system guarantees that if no new updates are made to the object, eventually all accesses will return the last updated value \cite{EV}. That means any history is acceptable under this consistency model, as we can extend any history to one that all $GET$ operations read the value of the last $PUT$ operation. Although this model provides maximum concurrency and availability, it makes the semantics of the shared memory complicated.

Causal consistency \cite{ahmed} is an intermediate consistency model which prevents some histories from occurring. Causal consistency is achievable with availability and partition-tolerance. We denote causal order by $\rightarrow$, and write $op_1 \rightarrow op_2$ if one of the following three conditions holds: 
\begin{enumerate}
\item $op_1$ and $op_2$ are two operations of a single process, and $op_1$ is before $op_2$ in the program order.  
\item $op_1$ is a $PUT$ operation, $op_2$ is a $GET$ operation, and $op_2$ reads the value written by $op_1$. 
\item There is some other operation $op_3$ such that $op_1 \rightarrow op_3$, and  $op_3 \rightarrow op_2$.
\end{enumerate}

Let $H_{i+p}$ comprises all operations of process $p_i$ and all $PUT$ operations in  $H$. Now we define causally consistent history as follows:

\begin{definition} [Causally Consistent History]
\label{def:cau}
A history $H$ is causally consistent if it can be extended (by appending zero or more response events) to some history $H'$ such that for each process $p_i$: 
\begin{itemize}
\item $complete(H'_{i+p})$ is equivalent to some legal sequential history $S$, and  
\item $S$ respects $\rightarrow$.
\end{itemize}
\end{definition}

\begin{figure}

\begin{center}
\includegraphics[scale=0.9]{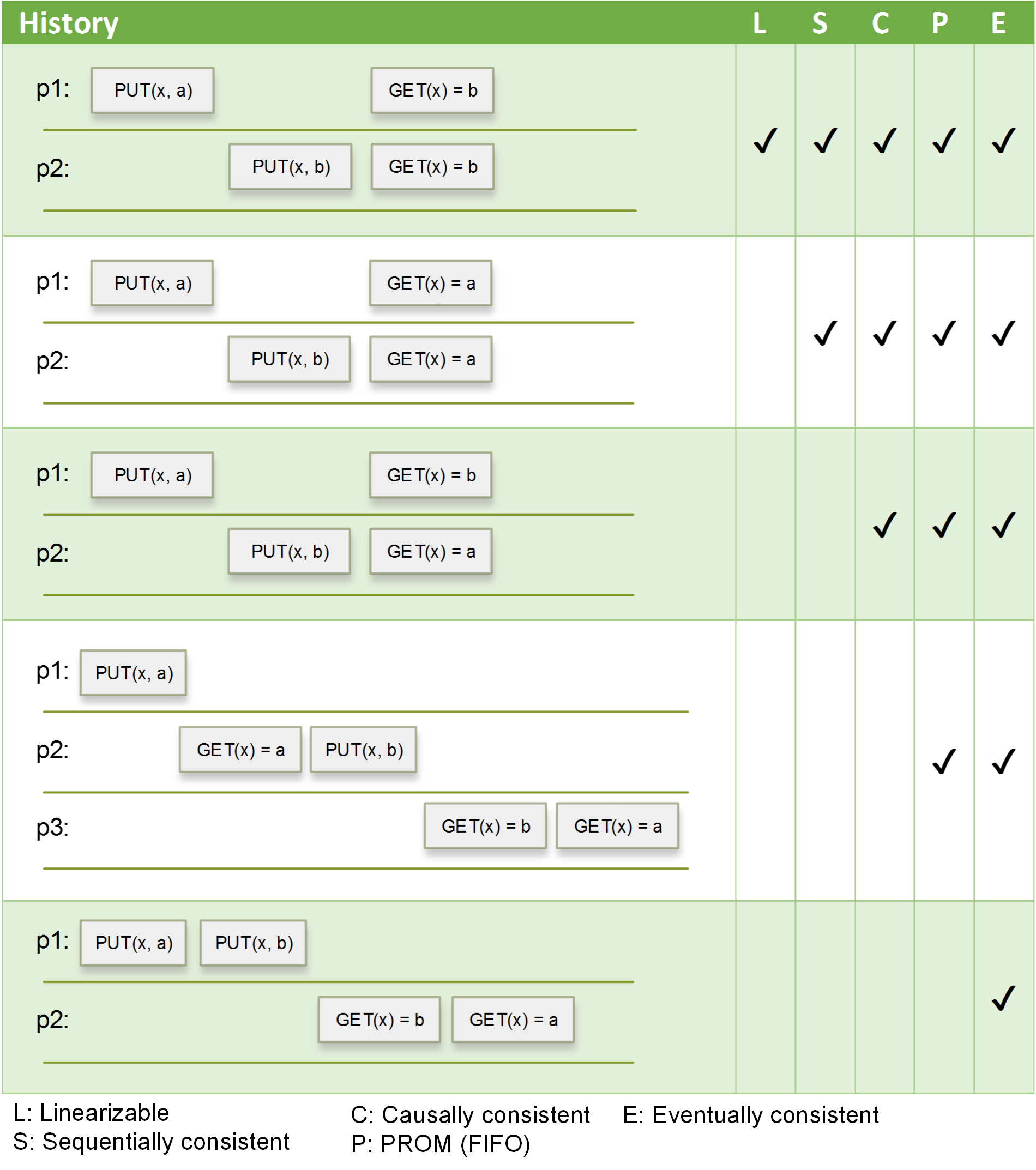}
\caption{Examples histories and consistency models}
\label{fig:Table}
\end{center}

\end{figure}

If we remove the second condition in Definition \ref{def:cau}, we have the definition of PRAM (FIFO) consistency model \cite{ahmed}. Thus, PRAM only respects programs orders. Figure \ref{fig:Table} provides some example histories along with the list of consistency models under which each history is allowed. 

\section{Causal Consistency in Distributed Data Stores}
\label{sec:CMD}
In distributed data stores, data objects are usually multi-version and each version specifies a value for an object. We say that version $X$ of object $x$ is visible for a node if the node has $X$ in its version chain for object $x$. Version $X$ of object $x$ causally depends on version $Y$ of object $y$ if the $PUT(y, Y) \rightarrow PUT(x, X)$. In that case, we also say $Y$ is an dependency for $X$. A distributed data store is causally consistent if when version $X$ is visible to a node, all of its dependencies are also visible.

\begin{figure}
\begin{center}
\includegraphics[scale=0.43]{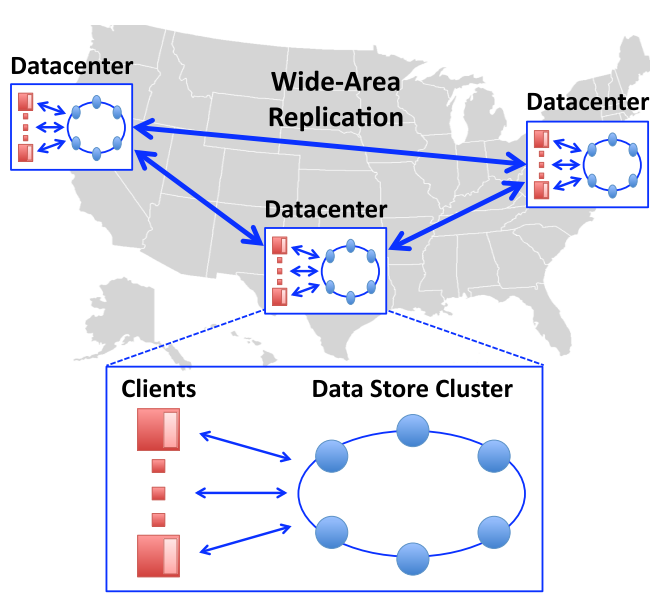}
\caption{The General Architecture of Distributed Data Stores \cite{COPS}.}
\label{fig:DS}
\end{center}
\end{figure}

COPS and GentleRain are two causally consistent data stores. Both COPS \cite{COPS} and GentleRain \cite{Gen} assume the same architecture for a distribute data store as shown in Figure \ref{fig:DS}. The distribute data store is a key-value store where each record is only a pair of a key and a value. As shown in Figure \ref{fig:DS}, the whole system consists of several geo-replicated datacenters each of which stores the whole key space. Each local datacenter provides linearizability for its local operations. Linearizability is achievable locally, since we assume there is no partition inside a datacenter. 

Both COPS and GentleRain assume that clients  use only their local datacenter. Thus, no client uses more than one datacenters. Whit this assumption and linearizability of local operations, we have the following observation: 

\begin{observation}
\label{obs:1}
Any version created in the datacenter d, may not be dependent on a version not visible in d.
\end{observation}

In order to provide a large key space, COPS and GentleRain provide partitioning of data across multiple nodes in each datacenter. Thus, each node in a datacenter is responsible for a range of keys. The node which is responsible for a key is called \textit{primary storage node} for that key.

\subsection{COPS}

COPS \cite{COPS} guarantees causal consistency by keeping the track of the dependencies between versions. In COPS, clients local to a datacenter perform their requests through a component called client library. Client library maintains a context for each client. A client context is a set of $\langle k, ver \rangle$ pairs showing the last version of objects seen by the client.

When a client preforms $GET(k)$, the client library forwards the operation to the primary storage node of $k$. This node simply reports the value $val$ and version $ver$ of the last version of object $k$ to the client library. The client library after reporting $val$ to the client, adds $\langle k, ver \rangle$ to the client's context. Adding  $\langle k, ver \rangle$ to the client's context indicates that this client has seen the version $ver$ of object $k$.

When a client performs $PUT(k, val)$, the client library forwards the operation along with the client's context to the primary storage node of $k$. This node assigns version number $ver$ to this value. The high-order bits in the version number are set to Lamport timestamp \cite{VC} and low-order bits are set to the node identifier. Primary storage node then writes the new version locally, and make it visible immediately. (According to Observation \ref{obs:1}, visibility of $ver$ respects causal consistency).

Primary storage node, next, sends $PUT\_{AFTER}(k, val, context, ver)$ messages to the primary storage nodes of $k$ in other datacenters. Each primary storage node upon receiving $PUT\_{AFTER}(k, val, context, ver)$, first ensures that all $\langle key, version\rangle$ pairs in the $context$ are visible in the datacenter. For each $\langle key, version\rangle$ pair not hosted on the primary storage node, it sends a $DEP\_{CHECK}(key,version)$ message to the node responsible for $key$. Each node upon receiving $DEP\_{CHECK}(key,version)$ responds immediately if $\langle key, version\rangle$ is visible, otherwise blocks. When the primary storage made sure that all dependencies are visible, makes version $ver$ visible.

\subsection{GentleRain}

GentleRain \cite{Gen} guarantees causal consistency by checking the timestamps of versions. It assigns each version a timestamp equal to the physical clock of the node. We denote the timestamp assigned to $X$ by $X.t$. GentleRain assigns timestamps such that following condition is satisfied:
\begin{itemize}
\item [I1:] If version $X$ of object $x$ depends on version $Y$ of object $y$, then $Y.t < X.t$.
\end{itemize} 
GentleRain associates every datacenter a variable called Global Stable Time (GST). For sake of space, the details of how each node computes GST is not provided in this paper, but basically, each node periodically computes GST such that following condition is satisfied: 
\begin{itemize}
\item [I2:] When GST in a node has a certain value $T$, then all versions with timestamps smaller than or equal to $T$ are visible in the datacenter.
\end{itemize}
When a client preforms $GET(k)$, the primary storage of object $k$, returns the newest version of $k$ which is either created locally, or has a timestamp no greater than GST. According to conditions I1 and I2 defined above, and Observation \ref{obs:1}, any version which is a dependency of $ver$ is visible in the local datacenter.

In order to perform $PUT(k, val)$, client sends the timestamp $t$ of the last version seen by the client together with the $PUT$ operation. The primary storage of object $k$, first waits until its physical clock goes higher than $t$, and then creates the new version. This wait is necessary to guarantee condition I1.

\subsection{Comparison of COPS and GentleRain}

In this section we compare COPS and GentleRain from two different aspects including throughput and Update Visibility Latency (UVL). 

\textit{\textbf{Throughput}}. We define throughput as the number of client operations per second. Each server has a limited message processing capability. That means, each server can process a limited number of messages per second. In COPS, part of this capability is allocated to process dependency check messages between partitions. Sharing message processing capability of servers between client operations and dependency check messages reduces the overall throughput of the COPS for client operations. On the other hand, since GentleRain does not send any dependency check message, its throughput is as good as eventual consistency where we do not check any dependency, and each version becomes visible once it reaches a replica. 

\begin{figure}
\begin{center}
\includegraphics[scale=0.33]{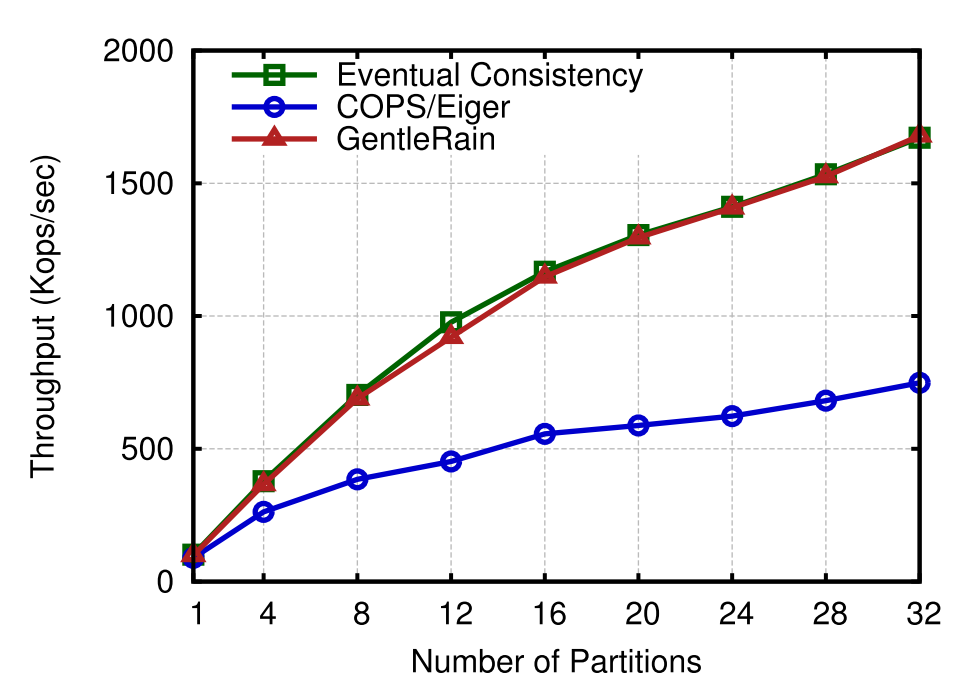}
\caption{Throughput of 1 to 32 partitions. A client reads an item from every partition  and updates an item at one partition \cite{Gen}. }
\label{fig:CGT}
\end{center}
\end{figure}

Figure \ref{fig:CGT} compares the throughput of GentleRain, COPS, and eventual consistency in an experiment where each client reads a randomly selected item from every partition and updates a randomly selected item at one partition. As it is shows in the Figure \ref{fig:CGT}, the gap between throughput of GentleRain and COPS grows as the number of partition grows. It is due to the fact that when number of partitions grows, average number of dependency check messages grows as well. 

\begin{figure}
\begin{center}
\includegraphics[scale=0.33]{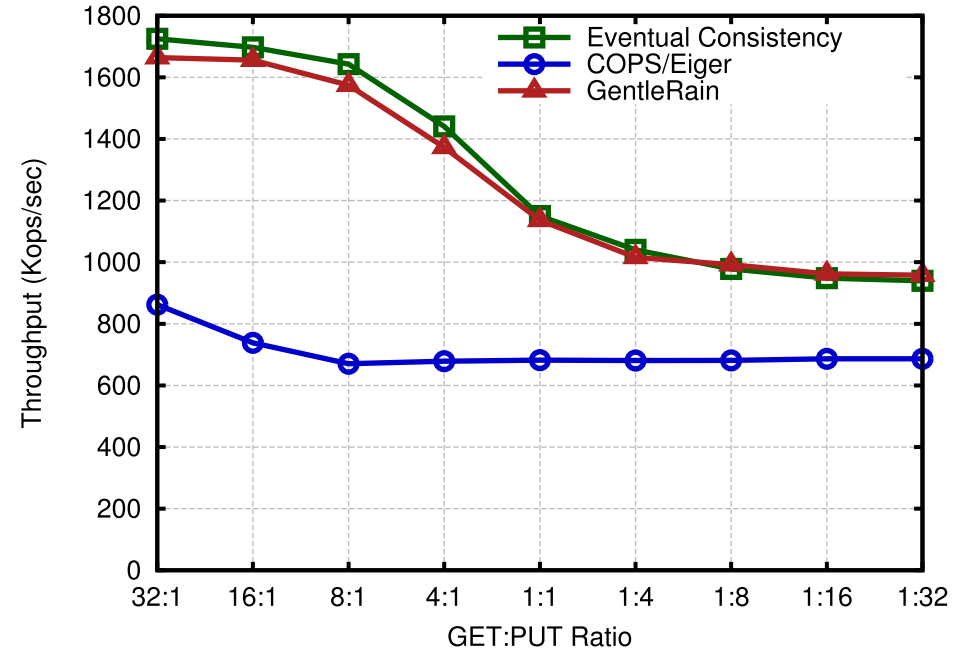}
\caption{Throughput with different GET:PUT ratios \cite{Gen}. }
\label{fig:CGT2}
\end{center}
\end{figure}

Figure \ref{fig:CGT2} shows the result of another experiment where each client reads $N$ randomly selected items from randomly selected partitions and write one randomly selected item to each of $M$ randomly selected partitions. Figure \ref{fig:CGT2} shows that the gap between GentleRain and COPS decreases as we move toward write-heavy end. That is due to the fact that as the number of $GET$ operations decrease, client contexts become lighter. Thus, COPS no longer need to check many dependencies for updates.

\textit{\textbf{UVL}}. We define UVL for an update as the time interval between its visibility in its original datacenter and its visibility in a remote datacenter. 
The mechanism of $GET$ operation in GentleRain may lead to larger UVLs comparing to COPS. In order to illustrate, consider the following example: Alice performs a $PUT$ operation on object $x$ at time $t$ before reading any other object. The $PUT$ operation creates version $X$. Since Alice had not read any object before the $PUT$ operation, $X$ does not dependent on any other version. Thus, $X$ can immediately be visible in any datacenter without violation of causal consistency. COPS does not send any dependency check message for this $PUT$ operation, and $X$ immediately becomes visible. On the other hand, GentleRain delays the visibility of $X$ in other datacenters until GST becomes equal to or greater than $X.t$.

In COPS, UVL is roughly equal to the Network Travel Time (NTT) between original datacenter and the remote datacenter. On the other hand, in GentleRain UVL is roughly equal to longest NTT in the whole system \cite{Gen}.

\begin{figure}
\begin{center}
\includegraphics[scale=0.8]{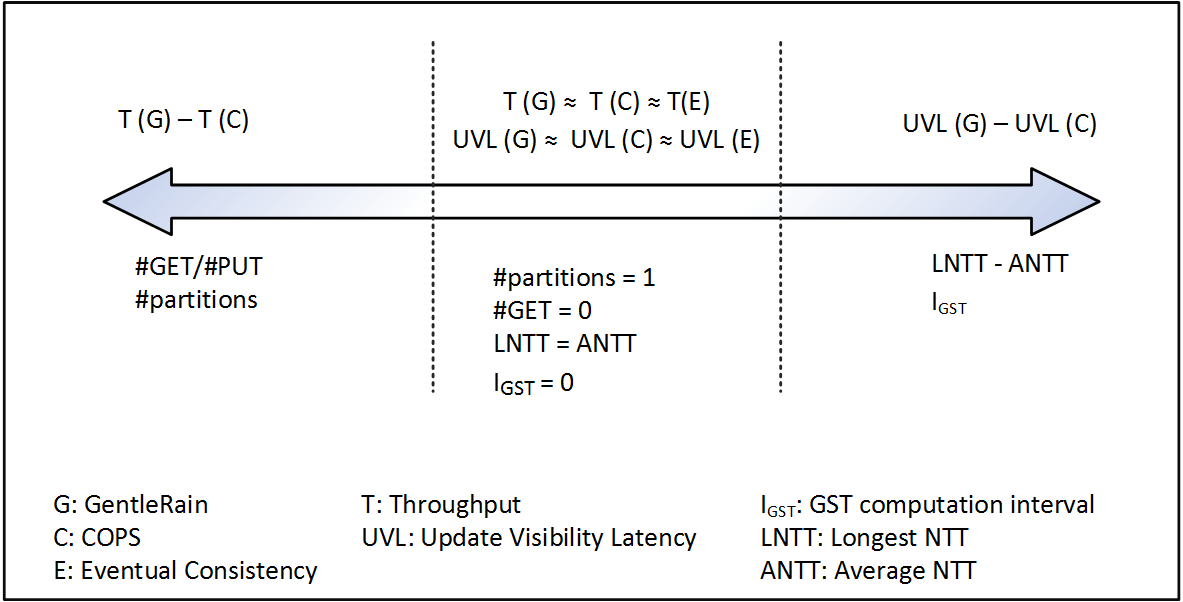}
\caption{Comparison of COPS and GentleRain.}
\label{fig:CGcom}
\end{center}
\end{figure}

Based on discussion above, Figure \ref{fig:CGcom} shows which system is better for which scenario. In the middle of the Figure \ref{fig:CGcom}, where: 1) we have only one partition, 2) only PUT operations are performed, 3) longest NTT equals average NTT, and 4) GST computation interval is 0, both COPS and GentleRain perform like eventual consistency regarding throughput and UVL. 

Right arrow shows that regarding UVL, COPS is better than GentleRain, and the gap between UVL of COPS and GentleRain increases as the difference between longest NTT and average NTT increases, or GST computation interval of GentleRain becomes longer. It shows that if we concern UVL more than throughput, then COPS is the better choice, especially if the longest NTT is large in the system.

Left arrow shows that regarding throughput, GentleRain is better than COPS, and the gap between throughput of GentleRain and COPS increases as the ratio of $GET$ operations to $PUT$ operations increases, or number of partition grows. It shows that if we concern throughput more than UVL, GentleRain is the better choice, especially for read-heavy workloads.  

\section{Dynamo}
\label{sec:Dy}

Dynamo is a distributed key-value store built for Amazon's platform. Amazon uses Dynamo to manage the state of some of its services with high availability requirements\footnote{Dynamo now is also publicly available under the name DynamoDB.}. Dynamo replicates each data object on $N$ nodes. Each key is assigned to a coordinator node which is one of the $N$ nodes storing the key. The list of nodes that can be used for storing a particular key is called the \textit{preference list}. The first node in the preference list is the key coordinator. The first $N$ nodes of the preference list are the $N$ nodes mentioned above. We refer to these $N$ nodes as top $N$ nodes. In addition to these top $N$ nodes, the preference list includes some other nodes to provide more availability in case of network partitions and node failures. Each $GET$ or $PUT$ operation involves the first $N$ \textit{healthy and reachable} nodes in the preference list. Thus, if some of top $N$ nodes are down or unreachable, we use lower-ranked nodes in the preference list.

When a client sends a $PUT(k, val)$ request, the coordinator creates a new version for $k$. Dynamo uses vector clock \cite{VC} timestamps as version numbers. The coordinator, after writing locally, sends the new version to $N$ highest-ranked healthy nodes in the preference list of $k$. If at least $W-1$ nodes respond, then the write is considered successful.

Similarly, when a client sends a $GET(k)$ request, the coordinator asks $N$ highest-ranked healthy nodes in the preference list of $k$. The coordinator then waits for $R$ responses. It may receive different versions of $k$. In that situation, two cases are possible: 1) All versions are causally related. In that case, the coordinator reports the most recent version regarding the vector clock timestamps. 2) Some of the versions are not causally related. In that case, the coordinator reports all causally unrelated versions to the client.

One of the advantages of Dynamo is that application designers who use Dynamo can configure it to achieve their desired levels of consistency and availability. Basically, achieving higher availability is possible by assigning lower values for $R$ and $W$. Some services such as shopping cart should be "always-writable", because rejection of writes to shopping cart adversely affects customers' satisfaction. Setting $W$ to 1 ensures that a $PUT$ operation is never rejected as long as at least one node in the system has written the new version value. However, it may increase the risk of inconsistency. In contrast to shopping cart, some services such as services that maintain product catalogs and promotional items, have high number of read requests and small number of updates. For such services, typically $R$ is set to 1 and $W$ to $N$. The typical $\langle N, R, W \rangle$ configuration used in Dynamo instances is $\langle 3, 2, 2 \rangle$ \cite{Dy}. 

\section{Comparison of Dynamo with COPS and GentleRain}
\label{sec:D_CG_com}

In this section we compare Dynamo with COPS and GentleRain regarding different aspects:

\textit{\textbf{Ability to tradeoff between availability and consistency}}. As described in Section \ref{sec:Dy}, Dynamo allows application designers to tradeoff between availability and consistency by configuring $N$, $R$, and $W$. On the other hand, both COPS and GentleRain enforce causal consistency, and do not provide any control to tradeoff between availability and consistency.

\textit{\textbf{Partition-tolerance}}. All three methods can provide availability in presence of network partitions. However, certain situations can make Dynamo unavailable. In particular, if there are not $W$ (respectively $R$) healthy and reachable nodes in the preference list of a key, $PUT$ (respectively $GET$) operations on that key become unavailable.

\textit{\textbf{Cost of achieving causal consistency}}. Dynamo can guarantee causal consistency by a proper configuration. In particular, if we limit preference list of each key to only top $N$ nodes, and set $R$ and $W$ such that $R+W>N$, then causal consistency is guaranteed. However, this configuration leads to long latencies and even unavailability in case of network partitions. On the other hand, COPS and GentleRain provide causal consistency while they are available in case of network partitions. In addition, the latencies of operations are much lower than latencies of Dynamo in the aforementioned configuration, as in COPS and GentleRain the $PUT$ and $GET$ operations do not need to wait for other replicas.

\textit{\textbf{Datacenter failure}}. In case of a datacenter failure, both COPS and GentleRain loss all updates originated in the failed datacenter not copied to other datacenters. On the other hand, in the Dynamo by setting $W$ to be more than 1 and by distributing replicas among different datacenters, we can decrease the risk of losing data.

\textit{\textbf{Locality of traffic}}. As mention in Section \ref{sec:CMD}, both COPS and GentleRain assume that clients local to a datacenter, exclusively use their local datacenter. That assumption is necessary to guarantee causal consistency by methods used in COPS and GentleRain. However, the locality of traffic equals to unavailability in case of datacenter failure for clients local to the failed datacenter. On the other hand, in Dynamo where clients can use any node to perform an operation, failure of a datacenter does not make system unavailable.

\textit{\textbf{Inter-replica communications}}. In COPS and GentleRain, $GET$ operations do not need to any communication with other replicas. On the other hand, in Dynamo we need to ask all replicas and wait for at least $R$ responses. For $PUT$ operations, although in all three systems, we need to replicate the updated value to all replicas, Dynamo needs more communication to finalize a $PUT$ operation. 

\textit{\textbf{Throughput}}. In \cite{Gen}, authors have compared the throughput of GentleRain and COPS with the throughput of an eventually consistent data store. In all cases, the throughput of the eventually consistent data store is higher than the throughput of GentleRain and COPS. Although Dynamo is an eventually consistent data store, we believe it differs from eventually consistent data store considered in \cite{Gen}. In particular, in Dynamo if $R$ (or $W$) is set to be greater than 1, for each $GET$ (or $PUT$) operation, we need to wait for the network travel time, but eventually consistent data store considered in \cite{Gen} does no include this wait. Therefore, we suggest additional experiments to compare the throughput of Dynamo with the throughput of GentleRain and COPS.

\textit{\textbf{Update visibility latency}}. We discussed  update visibility latency for COPS and GentleRain in Section \ref{sec:CMD}. We defined visibility latency for an update as the time interval between its visibility in its original datacenter and its visibility in a remote datacenter. In Dynamo, visibility latency depends on different factors including values of $R$ and $W$. However, we can say that there is no necessary visibility latency in Dynamo due to consistency check.

\section{Conclusion and Future Research}
\label{sec:Con}
In this paper, we focused on the problem of consistency in distributed data stores. We defined consistency in different levels. As the consistency is stronger, the semantics of the data store is simpler. However, achieving stronger consistency is more costly in terms of the availability. In particular, strong consistency provides a simple semantics for the application programmers, but according to the CAP theorem it is impossible to guarantee strong consistency for always available systems in presence of network partitions. In order to address this impossibility result, weaker consistency models are proposed including causal and eventual consistency models.

We reviewed two causally consistent data stores namely COPS and GentleRain. COPS guarantees causal consistency by explicit dependency check via dependency check messages. We explained how dependency check messages can adversely affect system throughput in COPS. On the other hand, GentleRain guarantees causal consistency by implicit dependency check via physical clock timestamps. GentleRain provides better throughput than COPS, but it may increase the update visibility latency. We believe COPS is more suitable for write-heavy workloads with relatively low dependencies. On other hand, GentleRain is more suitable for high throughput read-heavy workloads with large number of dependencies.

We also reviewed Dynamo which is an eventually consistent data store used by Amazon. An important advantage of Dynamo is that clients can configure it to achieve different levels of availability and consistency. This flexibility makes Dynamo more suitable than COPS and GentleRain for systems where clients need tight control over the tradeoff between availability and consistency. However, if clients only need causal consistency, we anticipate that COPS and GentleRain provide better availability and throughput than Dynamo. We suggest additional experiments to compare the throughput of Dynamo with GentleRaind and COPS.

A disadvantage of GentleRain is the delay in $PUT$ operations which is necessary to ensure that if version $X$ depends on version $Y$,then $Y.t < X.t$. We can eliminate this delay by using Hybrid Logical Clocks (HLC) \cite{HLC}. HLC is a clock that provides timesatmps close to physical clock, but unlike physical clock, satisfies logical clock condition ($e \rightarrow r \implies e.t_{hlc} < f.t_{hcl}$). The throughput improvement obtained by using HLC for GentleRain needs to be investigated. 

Another disadvantage of GentleRain (and COPS) is the loss of new updates in case of datacenter failures. In order to alleviate this problem, and to increase the durability of data, we can apply the method used in Dynamo for $PUT$ operations which is waiting for $W$ responses from other datacenters. The impact of this method on the throughput also needs to be investigated.

\bibliographystyle{unsrt}
\bibliography{bib}

\begin{thebibliography}{10}

\bibitem{ahmed}
Mustaque Ahamad, Gil Neiger, JamesE. Burns, Prince Kohli, and PhillipW. Hutto.
\newblock Causal memory: definitions, implementation, and programming.
\newblock {\em Distributed Computing}, 9(1):37--49, 1995.

\bibitem{Dy}
Giuseppe DeCandia, Deniz Hastorun, Madan Jampani, Gunavardhan Kakulapati,
  Avinash Lakshman, Alex Pilchin, Swaminathan Sivasubramanian, Peter Vosshall,
  and Werner Vogels.
\newblock Dynamo: Amazon's highly available key-value store.
\newblock {\em SIGOPS Oper. Syst. Rev.}, 41(6):205--220, October 2007.

\bibitem{Gen}
Jiaqing Du, C\u{a}lin Iorgulescu, Amitabha Roy, and Willy Zwaenepoel.
\newblock Gentlerain: Cheap and scalable causal consistency with physical
  clocks.
\newblock In {\em Proceedings of the ACM Symposium on Cloud Computing}, SOCC
  '14, pages 4:1--4:13, New York, NY, USA, 2014. ACM.

\bibitem{CAP}
Seth Gilbert and Nancy Lynch.
\newblock Brewer's conjecture and the feasibility of consistent, available,
  partition-tolerant web services.
\newblock {\em SIGACT News}, 33(2):51--59, June 2002.

\bibitem{Lin}
Maurice~P. Herlihy and Jeannette~M. Wing.
\newblock Linearizability: A correctness condition for concurrent objects.
\newblock {\em ACM Trans. Program. Lang. Syst.}, 12(3):463--492, July 1990.

\bibitem{HLC}
SandeepS. Kulkarni, Murat Demirbas, Deepak Madappa, Bharadwaj Avva, and Marcelo
  Leone.
\newblock Logical physical clocks.
\newblock In MarcosK. Aguilera, Leonardo Querzoni, and Marc Shapiro, editors,
  {\em Principles of Distributed Systems}, volume 8878 of {\em Lecture Notes in
  Computer Science}, pages 17--32. Springer International Publishing, 2014.

\bibitem{Seq}
L.~Lamport.
\newblock How to make a multiprocessor computer that correctly executes
  multiprocess programs.
\newblock {\em Computers, IEEE Transactions on}, C-28(9):690--691, Sept 1979.

\bibitem{VC}
Leslie Lamport.
\newblock Time, clocks, and the ordering of events in a distributed system.
\newblock {\em Commun. ACM}, 21(7):558--565, July 1978.

\bibitem{COPS}
Wyatt Lloyd, Michael~J. Freedman, Michael Kaminsky, and David~G. Andersen.
\newblock Don't settle for eventual: Scalable causal consistency for wide-area
  storage with cops.
\newblock In {\em Proceedings of the Twenty-Third ACM Symposium on Operating
  Systems Principles}, SOSP '11, pages 401--416, New York, NY, USA, 2011. ACM.

\bibitem{EV}
Werner Vogels.
\newblock Eventually consistent.
\newblock {\em Commun. ACM}, 52(1):40--44, January 2009.

\end{thebibliography}
\end{document}